\documentclass[conference]{IEEEtran}
\pdfoutput=1
\makeatletter

\def\ps@IEEEtitlepagestyle{%
  \def\@oddfoot{\mycopyrightnotice}%
  \def\@evenfoot{}%
}
\def\mycopyrightnotice{%
  {\footnotesize XXX-X-XXXX-XXXX-X/XX/\$XX.00~\copyright~20XX IEEE\hfill}
  \gdef\mycopyrightnotice{}
}

\usepackage{blindtext}
\usepackage{eso-pic}
\IEEEoverridecommandlockouts
\usepackage{cite}
\usepackage{amsmath,amssymb,amsfonts}
\usepackage{algorithmic}
\usepackage{graphicx}
\usepackage{textcomp}
\usepackage{xcolor}
\usepackage{algorithm}
\usepackage{array}
\usepackage[caption=false,font=normalsize,labelfont=sf,textfont=sf]{subfig}
\usepackage{stfloats}
\usepackage{url}
\usepackage{verbatim}
\usepackage{lipsum}
\usepackage{hyperref}
\usepackage{multirow}
\usepackage{listings}
\usepackage{tabularx}
\usepackage{booktabs}
\hypersetup{
    allcolors=black,
    colorlinks=true,
    linkcolor=black,
    urlcolor=blue,
}
\usepackage{tikz}

\def\BibTeX{{\rm B\kern-.05em{\sc i\kern-.025em b}\kern-.08em
    T\kern-.1667em\lower.7ex\hbox{E}\kern-.125emX}}
    
\usepackage{eso-pic}
\newcommand\AtPageUpperMyright[1]{\AtPageUpperLeft{%
 \put(\LenToUnit{0.17\paperwidth},\LenToUnit{-2cm}){%
     \parbox{0.9\textwidth}{\raggedleft\fontsize{8}{11}\selectfont #1}}%
 }}%
\newcommand{\conf}[1]{%
\AddToShipoutPictureBG*{%
\AtPageUpperMyright{#1}
}
}

\definecolor{codegreen}{rgb}{0,0.6,0}
\definecolor{codegray}{rgb}{0.5,0.5,0.5}
\definecolor{codepurple}{rgb}{0.58,0,0.82}
\definecolor{backcolour}{rgb}{0.95,0.95,0.92}

\lstdefinestyle{mystyle}{
  backgroundcolor=\color{backcolour},   commentstyle=\color{codegreen},
  keywordstyle=\color{magenta},
  numberstyle=\tiny\color{codegray},
  stringstyle=\color{codepurple},
  basicstyle=\ttfamily\footnotesize,
  breakatwhitespace=false,         
  breaklines=true,                 
  captionpos=b,                    
  keepspaces=true,                 
  numbers=left,                    
  numbersep=5pt,                  
  showspaces=false,                
  showstringspaces=false,
  showtabs=false,                  
  tabsize=2
}

\newcommand{\orcidicon}{\includegraphics[width=0.32cm]{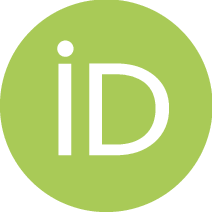}}

\foreach \x in {A, ..., Z}{%
\expandafter\xdef\csname orcid\x\endcsname{\noexpand\href{https://orcid.org/\csname orcidauthor\x\endcsname}{\noexpand\orcidicon}}
}

\begin{document}
\title{\vspace*{1cm} Substation Bill of Materials: A Novel Approach to Managing Supply Chain Cyber-risks on IEC 61850 Digital Substations\\
{}
\thanks{This work has been financed by the Department of Economic Development, Sustainability, and Environment of the Basque Government under the ELKARTEK 2023 program, project BEACON (with registration number KK-202300085).}}

\author{
\IEEEauthorblockN{1\textsuperscript{st} Xabier Yurrebaso$^{1}$~\orcidA{}, 2\textsuperscript{nd} Fernando Ibañez$^{2}$~\orcidB{}, 3\textsuperscript{rd} \'Angel Longueira-Romero$^{1}$~\orcidC{}}
\IEEEauthorblockA{\textit{$^{1}$Digital - CORES, $^{2}$TECU - Grids and Energy System} \\
\textit{TECNALIA, Basque Research and Technology Alliance (BRTA)}\\
Derio, Spain \\
\{xabier.yurrebaso, fernando.ibanez, angel.longueira\}@tecnalia.com}

}


\maketitle
\conf{\textit{  V. International Conference on Electrical, Computer and Energy Technologies (ICECET 2025) \\ 
3-6 July 2025, Paris-France}}

\begin{abstract}
    Smart grids have undergone a profound digitization process, integrating new data-driven control and supervision techniques, resulting in modern digital substations (DS). Attackers are more focused on attacking the supply chain of the DS, as they a comprise a multivendor environment. In this research work, we present the Substation Bill of Materials (Subs-BOM) schema, based on the CycloneDX specification, that is capable of modeling all the IEDs in a DS and their relationships from a cybersecurity perspective. The proposed Subs-BOM allows one to make informed decisions about cyber risks related to the supply chain, and enables managing multiple DS at the same time. This provides energy utilities with an accurate and complete inventory of the devices, the firmware they are running, and the services that are deployed into the DS. The Subs-BOM is generated using the \textit{Substation Configuration Description} (SCD) file specified in the IEC 61850 standard as its main source of information.
    
    We validated the Subs-BOM schema against the Dependency-Track software by OWASP. This validation proved that the schema is correctly recognized by CycloneDX-compatible tools. Moreover, the Dependency-Track software could track existing vulnerabilities in the IEDs represented by the Subs-BOM.
\end{abstract}

\section{Introduction}
    \IEEEPARstart{I}{n} recent years, smart grids have undergone a profound process of digitization. Until recently, the facilities responsible for the generation and distribution of energy were considered isolated elements of the grid. Therefore, it was sufficient to adopt physical security measures for the protection of the equipment deployed on them. However, the evolution of Information and Communication Technologies (ICT) has increased the monitoring and remote control capabilities of the Smart Grid infrastructure. New data-driven control and supervision are being developed to solve identified challenges in the management of the smart grid. Distribution System Operators (DSOs), aware of the need for improvements in grid monitoring and control, are beginning the transition towards the deployment of new modern Digital Substations (DS). In the coming years, it is estimated that the deployment of DS will surpass that of traditional substations worldwide~\cite{substation_Market_Size}.
    
    In this scenario, the IEC 61850 standard was created with the aim of systematizing the deployment of DS by standardizing all the communications that are involved in their operation. This international standard defines the communication protocols for Intelligent Electronic Devices (IEDs) at electrical substations.
    
    Nevertheless, the IEC 61850 standard by itself is not enough: the digitalization of the Smart Grid substations requires deploying and integrating a massive number of devices manufactured by several vendors with new digital services and capabilities. One concern that is arising worldwide is how DSOs will manage cyber-risks associated to these new multi-vendor DS. In other words, DSOs need to create processes and tools for the identification and assessment of cyber-risks posed by digital products in charge of the correct transportation and distribution of energy.

    In this multi-vendor scenario, attackers have directed their efforts towards the device supply chain, rather than directly attacking DS. This fact is supported by many national cybersecurity agencies, such as the European Union Agency for Cybersecurity (ENISA)~\cite{enisa_2030_cibersecurityChallenges}.

    One of the biggest and most relevant attacks that exploits the supply chain is the SolarWinds hack~\cite{solarWindsAttack_1}. SolarWinds mainly develops software for large business companies such as Microsoft, and CISCO. It also provides software services for US government institutions and agencies such as the Pentagon, Homeland Security, and the National Nuclear Security Administration. In this scenario, the attackers managed to affect the supply chain in a software delivered by SolarWinds known as Orion~\cite{solarWindsAttack_3} by inserting functions in the source code of a particular dynamic library file (.dll). Subsequently, SolarWinds signed and distributed the .dll as part of its update processes, infecting more than 18,000 customers and 40 public entities from different sectors with malware~\cite{solarWindsAttack_2}. This attack highlights the importance of managing the risks in the supply chain, particularly in critical applications and environments.
    
    In this research work, we present the Subs\nobreakdash-BOM schema, a novel approach that helps DSOs managing Supply Chain Cyber-Risks of IEC 61850 DS:

    \begin{itemize}
        \item The Subs-BOM schema is compliant with the IEC 61850 standard.
        \item It can be integrated in the engineering process described in the IEC 61850.
        \item The Subs-BOM format is based on widely adopted international standards such as CycloneDX.
        \item It allows to make informed decisions about the supply-chain cyber risks associated to a set of DS.
        \item The Subs-BOM schema enables managing multiple DS at the same time.
    \end{itemize}

    The solution proposed in this paper takes as input the \textit{Substation Configuration Description} (SCD) file specified in the IEC 61850 standard to generate the \textit{Substation Bill of Materials} (Subs-BOM) during the engineering process. The Subs-BOM schema proposed in this paper provides DSOs with an accurate and complete inventory of the devices, associated software and services that are deployed into their substations. The Subs-BOM contributes to a rapid identification of vulnerable components deployed into their Operational Technologies (OT) environments. This data can be used to trigger an incident response process to analyze, triage and remediate the detected vulnerabilities. Moreover, the Subs-BOM can be reused or extended to other domains within the power utility automation beyond the DS domain, e.g., modeling of hydropower plants, wind turbines or even substation to substation communications.

    The content of this papers is as follows: In Section~\ref{sec:background}, we described the main concepts related to the supply chain attacks and the IEC 61850 standard. In Section~\ref{sec:relatedWork}, we reviewed previous research works targeting supply chain attacks in DS. The proposed Subs-BOM schema is defined in Section~\ref{sec:proposedApproach}, and then, validated in Section~\ref{experimental_setup}. Finally, the results are discussed in Section~\ref{sec:resultsDiscussion}, and the conclusions of this research are presented in Section~\ref{sec:conclusions}.
\section{Background}
\label{sec:background}
In this section, we introduce the main concepts related to risk management in digital substations: (1) Supply chain attacks, (2) Bill of Materials (BOM), and (3) the IEC 61850 Standard.

\subsection{Supply Chain Attack}
    According to the European Repository of Cyber Incidents (EuRepoC), the energy sector is the fifth in terms of the number of cyber attacks against critical infrastructures~\cite{eu_repo_ci_database}. In their reports, the EuRepoC identifies Supply Chain Compromise as the main initial access techniques to DS~\cite{mitre_supply_chain_compro}.
    
    
    According to the European Union Agency for Cybersecurity (ENISA)~\cite{enisa_threatLandscape}, a \textit{Supply Chain Attack} involves first targeting the \textit{Software Development life cycle (SDLC)} of product vendors. So instead of attacking the device itself directly when it is deployed, this attack goes a step back, and compromises the source code that it runs in the early phases of development.
    
    Once the software of the product vendor is compromised, this attack targets the final customer that deploys the compromised product into their infrastructure. An attacker would have the ability to compromise a DS by compromising any of the software vendors involved in the development of any software component running in the final product.
    
    To deal with these attacks in DS, it is essential to establish a comprehensive management procedure that is able to identify the affected assets and provide a remediation response, both in a timely manner. This asset management is identified by the NIST as one of the main challenges faced by energy utilities in managing the cyber-risks of their OT infrastructures~\cite{nist_1800_23}. Nevertheless, many energy organizations rely on manual processes to manage their OT assets, which makes it challenging to quickly identify and respond to potential threats targeting DS~\cite{nist_1800_23}.
    
    Under this scenario, manufacturers should start working on the definition of processes to automate asset management. These efforts should focus towards the correct generation of the inventory of IEDs and auxiliary devices deployed on DS, as they are crucial to an exhaustive cybersecurity management.
\subsection{Bill of Materials (BOM)}
    The term \textit{Bill of Materials} is not new, especially in the world of electronics where BOMs have historically been used to define the different electronic components that make up a given product. This concept was quickly translated into other fields. In 2021, the US Executive Office of the President published the ```U.S. Executive Order 14028 - Improving the Nation's Cybersecurity'''~\cite{eo_improving_nations_cybersec} where the Software Bill Of Materials (SBOM) is described as ```a formal record containing the details and supply chain relationships of various components used in building software'''. In this sense, SBOMs try to perform the same function by being a complete inventory of the different software elements included in a product, enabling the analysis of licenses and vulnerabilities. This information can by further used to assess the risk of a product, and if combined with a machine-readable SBOM format, it allows automating the risk assessment process to a large extent.

    Due to the demonstrated usefulness of SBOM, it has been rapidly extended to other types of components and is not limited to software components alone. For example, Cryptographic Bill of Material (CBOM), Machine-learning Bill of Material (ML-BOM) or even Manufacturing Bill of Materials (MBOM). For this reason, it might be better to speak of eXtended Bill of Materials (xBOM) to refer to all the information that can be included about a given product to allow a quick and informed decision-making.

    \begin{table}[!htb]
    \caption{Minimum SBOM elements~\cite{ntia_min_sbom_req}.}
    \label{tab:min_sbom_elems}
    \begin{tabular}{| m{5em} | m{22em} |}
    \hline
     Data Fields & Document baseline information about each component that should be tracked: Supplier, Component Name, Version of the Component, Other Unique Identifiers, Dependency Relationship, Author of SBOM Data, and Timestamp.   \\ 
    \hline
     Automation Support & Support automation, including via automatic generation and machine-readability to allow for scaling across the software ecosystem \\ \hline
    
     Practices and Processes & Define the operations of SBOM requests, generation and use including: Frequency, Depth, Known Unknowns, Distribution and Delivery, Access Control, and Accommodation of Mistakes\\ \hline
    \end{tabular}
    \end{table}

    The Department of Commerce, in coordination with the National Telecommunications and Information Administration (NTIA), published which are the minimum elements that a SBOM should have to support the Executive Order (14028) on Improving the Nation’s Cybersecurity~\cite{ntia_min_sbom_req} (see Table~\ref{tab:min_sbom_elems}).
    
    Nowadays, there exist several SBOM formats in commercial use that are compliant with the specifications in Table~\ref{tab:min_sbom_elems}. Raihanul~\cite{sbom_analysis} presents a comprehensive study of SBOMs from Supply-chain Risk Management perspective. This research work concludes that the main SBOM formats used by the industry are (in no particular order): 

    \begin{itemize}
        \item Software Package Data eXchange (SPDX)~\cite{spdx}.
        \item CycloneDX~\cite{cyclone_dx}.
        \item SWID~\cite{SWID}.
    \end{itemize}

    Table~\ref{tab:sbom_format_comparison} shows a comparison between these formats.

\begin{table}[!htp]
\centering
\caption{SBOM format comparison.}
\label{tab:sbom_format_comparison}
\resizebox{\columnwidth}{!}{%
\begin{tabular}{@{}llll@{}}
\toprule
\textbf{SBOM Format}                  & \textbf{SPDX}                                                                         & \textbf{CycloneDX}                                            & \textbf{SWID} \\ \midrule
\textbf{Standardization Organization} & ISO/IEC                                                                               & ECMA                                                          & ISO/IEC       \\ \midrule
\textbf{Standardized Version}         & 2.2.1                                                                                 & 1.6                                                           & 2015          \\ \midrule
\textbf{Supported Formats}            & \begin{tabular}[c]{@{}l@{}}XML\\ JSON\\ RDF-XML\\ Tag-Value\\ YAML\\ XLS\end{tabular} & \begin{tabular}[c]{@{}l@{}}XML\\ JSON\\ Protobuf\end{tabular} & XML           \\ \midrule
\textbf{VEX / VDR Support}            & NO                                                                                    & YES                                                           & YES           \\ \bottomrule
\end{tabular}%
}
\end{table}
\subsection{The IEC 61850 standard}
\label{subsec:iec_61850_std}
    First published in 2003, the IEC 61850 standard describes the communication networks and systems for power utility automation~\cite{iec61850_standard}. Its initial scope was related to the standardization of communications between substation automation systems. Since then, it has evolved until becoming a central communication standard for both IEC TC 57 and the Smart Grid reference architectures~\cite{AFTAB2020106008}. Nowadays, the IEC 61850 standard has been updated and extended to include additional use cases. This standard has become a key driver in the the expansion of smart grid networks and the modernization of existing substations. It enables supporting the desired interoperability and flexibility in an ecosystem where multi-manufacturer solutions exist.

    Our research focuses on Part 6 of the IEC 61850 standard~\cite{iec_61850_part6} which defines:

    \begin{itemize}
        \item The \textbf{System Conﬁguration description Language (SCL)}, and all the types of SCL ﬁles required to describe the conﬁguration of a substation.
        \item The \textbf{SCL based engineering automation processes} to manage the substation assets.
    \end{itemize}

    \subsubsection{System Conﬁguration description Language (SCL)}
        The System Configuration description Language (SLC) is a language that allows to describe a model of a DS from the primary (power) system structure, the communication system, the application level communication, to each IED conforming the substation.

        Once the IEC 61850 defines the SCL, it sets the foundation to build different configuration files that upon the syntax specified by the SCL. Each configuration file is in charge of modeling a specific aspect of the DS (see Table~\ref{tab:scl_files}). More specifically, the System Configuration Description (SCD) file describes all the IEDs deployed in a DS.

        \begin{table}[!htb]
            \caption{Main SCL conﬁguration ﬁles defined by the IEC 61850 standard.}
            \label{tab:scl_files}
            \begin{tabular}{|c|m{7em}|m{17em}|}
                \hline
                 File & Name & Description \\ \hline 
                 SCD & System Conﬁguration Description  & It is the configuration by a system (substation) level conﬁgurator after processing all the ICD ﬁles from each IED and SSD ﬁles  \\ \hline
                 CID & Conﬁgured IED Description & Specific configuration of an IED for a specific DS  \\ \hline
            \end{tabular}
        \end{table}

        Any DS that is compliant with this standard must be modeled according to the SCL specifications, and therefore, must have an SCD file with all the information about its IEDs. Our proposal takes advantage of this fact to extract the necessary data to complete the Subs\nobreakdash-BOM model.

    \subsubsection{SCL based engineering automation processes}
        In addition to the SCL definition, the IEC 61850 Part 6 defines six different SCL file types (see table~\ref{tab:scl_files}). These files are generated or consumed by both IEDs and system configuration tools. The standard also specifies the roles that these tools shall play to be integrated in a SCL-based DS engineering process.
        
        More precisely, the standard defines the main functionalities expected for IED and system configuration tools regarding the SCL file management. For each tool, the IEC 61850 assigns specific capabilities that translate into (1) the generation, (2) editing or (3) processing of a SCL file or a part of it, for its own consumption or for another tool as part of the engineering process.
\section{Related Work}
\label{sec:relatedWork}
    In this research work, we propose a new approach to manage the cyber risks in the supply chain of DS using the Substation Bill of Material (Subs-BOM). Therefore, this section covers both risk management in the supply chain, and the integration of BOM-related technologies in the analysis.

    Liu et al.~\cite{rw_riskAssesmentSubstations_2015} propose a risk assessment procedure in terms of architecture and tripping logic.
    Lopez et al.~\cite{rw_intrusionDetection_2022} proposed a intrusion detection system that integrates the communication protocols defined in the IEC 61850 standard, together with the topology of the the substation. Liu et al.~\cite{rw_assetAnalysisRisk_2011} proposed an asset identification methodology further used for risk assessment of the DS. Khodabakhsh et al.~\cite{rw_cyberRisIdentification_2020} apply cyber-risk identification methodology with focus on DS for recognizing potential cyber-attacks, evaluates the cyber-risks and their impacts, and defines mitigation plans to ensure reliable operation of DS and safe and secure delivery of reliable power. These research works extract the information to model the substation from the SCL configuration file of the substation, but in their research, they do not propose any systematic data structure to store that data.

    Holik et al.~\cite{rw_threatModeling_2022} create a threat model of a digital secondary substation and the communications with the control center. This model combines simulation with emulated communication, and enables verification of threat likelihoods and impacts. Their results show that even publicly available tools can be easily used to disrupt grid communication and potentially cause loss of the entire grid's observability and controllability. Even though this research work reviews several threats for DS, the supply chain for the IEDs is not considered. This lack of considering the supply chain in threat analysis for DS is common in the literature~\cite{rw_intrusionDetectionSurvey_2021, AFTAB2020106008}.The authors also took advantage of the SCL language model defined in the IEC 61850. Nevertheless, they do not propose any approach to extract information in a systematic manner, or using a standardized data structure to process it.

    Duman et al.~\cite{rw_supplyChainDigitalSubstations_2024} propose a hardening system, namely, hardening framework for substations (HFS), to measure and optimally improve the security posture of substations against supply chain attacks. HFS provides a hardening mechanism for securing substations while considering the budget and operational constraints, together with a visual framework that allows operators to generate attack graphs and manually experiment with various hardening options. The authors validate the effectiveness of HFS based on several scenarios, showing that their approach improves the security postures of substations against supply chain attacks even with limited supply chain-related hardening options by reducing the number of successful supply chain attackers. Although this research work in really comprehensive, they do not specify how they obtain the data to generate the attack graph, or what information is needed to build it.
    Duman et al.~\cite{rw_modelingSupplyChainAttacks_2019} discuss the general concept and unique aspects of supply chain attacks. Then, they present concrete models of different supply chain attacks through extending the attack graph model and designing a security metric, namely k-Supply. Finally, the authors apply such models to quantitatively study the potential impact of supply chain attacks through simulations. Again, this research work neither describes which data is necessary to build the model, nor which is the source to obtained it.

    In general, research works that address risk assessment in DS lack the inclusion of supply chain attacks as a threat in their analysis. On the other hand, those which do consider supply chain attacks are obscure about what data is needed from the DS, or where it can be found. Finally, to the best of our knowledge, no research work proposed a standardized data structure that can model a DS from its components to the relations between them.

    Our proposal focuses on those premises for modeling a DS from a cybersecurity perspective: which data is needed for a comprehensive modeling, where the data can be found for all IEC 61850 compliant substations, and how to define a standardized data structure that is capable of representing all the elements and their relations.
\section{Proposed approach}
\label{sec:proposedApproach}
    DSOs need automated mechanisms to facilitate the management of OT assets deployed in their infrastructures, enabling effective management of cyber risks and threats. To perform such asset management, we propose the definition of a Substation Bill of Materials (Subs-BOM), whose generation could be fully automated during the Engineering process. The proposed Subs-BOM is based on the CycloneDX format for SBOMs, enriching its content using the information available in the SCD file of IEC 61850 compliant DS.

    In this section, we first select the SBOM format that we are going to use as the foundation for the Subs-BOM. Then, we define the schema for the Subs-BOM, its structure and fields. Finally, we demonstrate how the proposed Subs-BOM is built, how it can be integrated in the IEC 61850 engineering process, and we discuss some of its limitations.
    
    \subsection{CycloneDX as SBOM Format}
        In the industry, there exist multiple standards for SBOMs that are currently in use. The most popular ones are shown in Table \ref{tab:sbom_format_comparison}. Nevertheless, the Subs-BOM proposed in this paper is based on the CycloneDX format for the following reasons:

        \begin{enumerate}
            \item \textbf{CycloneDX is standardized}: It was designed in 2017 for use with OWASP Dependency-Track, an open-source component analysis platform that identifies risk in the software supply chain.

            \item \textbf{Integration with existing tools}: CycloneDX format is supported by most of the available tools in the market for supply chain risk management.

            \item \textbf{CycloneDX is not limited to software inventory}: It is a full-stack standard capable of recording different types of components, like software packages, firmware, device configuration, and even cryptographic keys.

            \item \textbf{CycloneDX is designed for native communication of cybersecurity}: This includes the Vulnerability Exploitability eXchange (VEX) or Vulnerability Disclosure Report (VDR).

            \item \textbf{CycloneDX is widely adopted}~\cite{2023_cylonedx_extended}: It is a format that is widely spread in the modern industry.
        \end{enumerate}

        These reasons make CycloneDX a good candidate to use as the foundation of the Subs-BOM.

    \subsection{Defining the fields in the Subs-BOM}
\begin{figure}[!htb]
            \centering
            \includegraphics[width=3in]{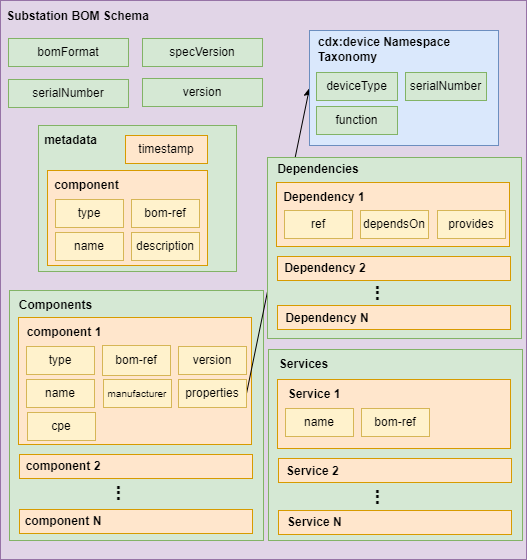}
            \caption{Proposed Substation BOM schema}
            \label{fig:subs_bom_schema}
        \end{figure}

        The proposed Subs-BOM schema is a subset of the CycloneDX v1.6 Reference specification~\cite{cyclonedx_16_spec}. In this section, we define the fields that comprise the Subs-BOM. Fig.~\ref{fig:subs_bom_schema} shows the proposed Subs-BOM schema.

        \subsubsection{General Information}
            The Subs-BOM is based on CyloneDX, so it also inherits most of its properties and fields. This means that the general information describing the Subs-BOM (e.g., format, version) is the same as in a SBOM defined in CycloneDX.6

        \subsubsection{Metadata Field}
            The metadata object contains the information that describes the main component of the Subs-BOM which is the IEC 61850 DS itself. CycloneDX does not support the definition of a \textit{substation} component type. For this reason, we extended the component types defined by CycloneDX Reference schema~\cite{cyclone_component_types} to include a new type called \textit{substation}. This proposed substation type serves as the foundation to define the remaining fields of the Subs-BOM.

        \subsubsection{Components Field}
            The \textit{substation} type can hold two types of CycloneDX components:

            \begin{itemize}
                \item \textbf{Device component type}: it describes each of the IEC 61850 IEDs identified in the SCD file of the substation. The information of each device is completed with some specific properties defined by the \textit{cdx:device Namespace Taxonomy} of the CycloneDX standard~\cite{cdx_device_taxonomy}.
            
                \item \textbf{Firmware component type}: As pointed out by the CycloneDX specification, when devices contain firmware, the BOM should also include an additional component of type \texttt{firmware} describing information about the software running on the device.
            \end{itemize}

            To group the data about the devices and their firmware, the Subs-BOMS defines the \textit{components} array. This field describes every component in which, directly or transitively, the substation depends on.

        \subsubsection{Services Field}
            Finally, the Subs-BOM defines the \textit{services} field. It contains the information of the several IEC 61850 services exposed by each IEDs (devices). The main purpose of the Subs-BOM is to enable and facilitate the risk management in DS. For this reason, it is critical to take into account these services as they expose data and functionalities of the IEDs to the DS process bus.

        \subsubsection{Dependencies Field}
            The Subs-BOM also defines the \textit{dependencies} field to group the information of the existing dependencies between the components (firmware and devices). This field allows generating the dependency graph of the substation.

    \subsection{Level of Detail of the Subs-BOM}
        The level of detail refers to how deep an element in the BOM is described, and therefore, it impacts directly in the complexity of the BOM. The higher the level of detail, the more the number of dependencies that have to be captured~\cite{tr_03183_german}.
        
\begin{figure}[!htb]
            \centering
            \includegraphics[width=3in]{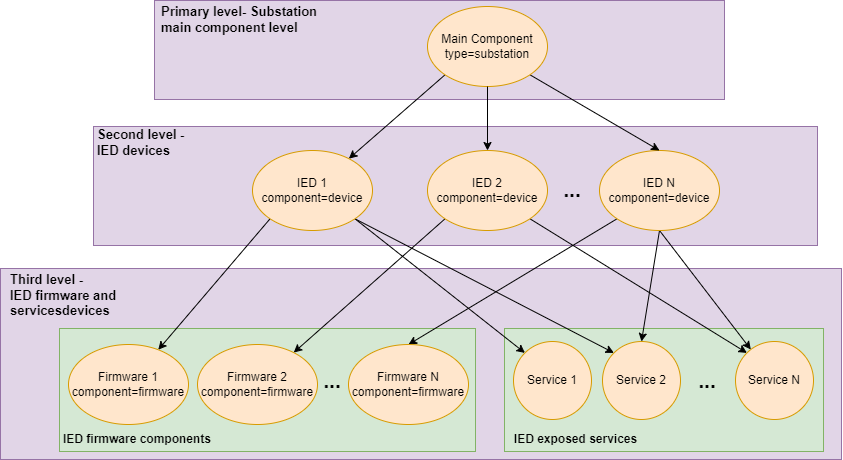}
            \caption{Proposed Substation BOM dependency graph}
            \label{fig:subs_bom_dep_graph}
        \end{figure}

        CycloneDX recommends one node of depth (two levels) to avoid problems with circular dependencies and other complex relationships between components~\cite{cyclone_dx}. However, CycloneDX also specifies that the firmware that runs in a device should also be included as an additional component in the SBOM~\cite{cyclonedx_16_spec}. Therefore, following this requirement, the Subs-BOM implements three levels of deepness to represent basic dependencies among components of the substation. This is shown in Fig.~\ref{fig:subs_bom_dep_graph}.

    \subsection{Generation Process of the Subs-BOM}
\begin{figure}[!htb]
            \centering
            \includegraphics[width=3in]{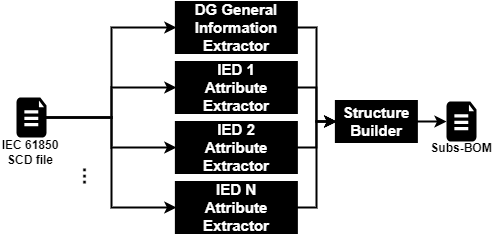}
            \caption{Proposed Substation BOM schema}
            \label{fig:subs_bom_generation}
        \end{figure}

        All the information needed to build the Subs-BOM is provided by the SCD file defined in the IEC 61850 standard. Fig.~\ref{fig:subs_bom_generation} shows the general building process of a Subs-BOM.

        The IEC 61850 standard provides a reference object model for a DS in Part 6~\cite{iec_61850_part6}. This model, divided into three main parts, serves as the source of information to build the proposed Subs-BOM:

        \begin{itemize}
            \item \textbf{Substation/Line/Process}: Information related to the substation, mainly its name, description, and topological information.

            \item \textbf{Product}: Information about all IEDs and Logical Node implementations (information model)~\cite{iec_61850_part7_1}.

            \item \textbf{Communication}: Communications-related objects and communication connections between IEDs.
        \end{itemize}

\begin{figure}[!htb]
            \centering
            \includegraphics[width=1in]{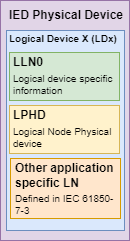}
            \caption{Physical and Logical device representation according to IEC 61850 standard }
            \label{fig:ied_modelization}
        \end{figure}

        In this reference object model defined by the IEC 61850, all the IED information is modeled as a \texttt{Product}. There, each IED is modeled as a physical device hosting logical nodes (specific task) and data. There is also information related to the real devices, and communication services. Fig.~\ref{fig:ied_modelization} shows the object model for an IED \texttt{Product} according to the IEC 61850. For building the Subs-BOM of a substation, we will only need to focus on the information included first two parts of the model: Substation/Line/Process and Product. This is because the Subs-BOM schema just models the IEDs in a DS and their Firmware (FW) dependencies.

        Every IED instance in the SCD file should include a Logical Node of type LPHD\cite{iec_61850_part7_3}. The information contained on LPHD comprises device nameplate of the physical device which hosts the Logical Device.

        In Table~\ref{tab:mapping_DPL}, we present the mapping between the attributes of the DPL data class associated to LPHD Logical Node, and the Subs-BOM of a component. DPL data class is fully described in Part 7-3 of the IEC 61850 standard \cite{iec_61850_part7_3}.
        
        \begin{table}[!htb]
            \centering
            \caption{Mapping of the Component Object.}
            \label{tab:mapping_DPL}
            \resizebox{\columnwidth}{!}{%
            \begin{tabular}{@{}lll@{}}
                \toprule
                \textbf{Subs-BOM ATTRIBUTE}              & \textbf{CycloneDX TYPE}   & \textbf{IEC 61850 ATTRIBUTE}        \\ \midrule
                \multirow{4}{*}{name}                    & device                    & IED.name                            \\ \cmidrule(l){2-3} 
                                                         & \multirow{3}{*}{firmware} & LPHD.PhyNam.vendor                  \\
                                                         &                           & LPHD.PhyNam.model                   \\
                                                         &                           & LPHD.PhyNam.swRev                   \\ \midrule
                \multirow{2}{*}{manufacturer}            & device                    & \multirow{2}{*}{LPHD.PhyNam.vendor} \\
                                                         & firmware                  &                                     \\ \midrule
                \multirow{2}{*}{version}                 & device                    & LPHD.PhyNam.hwRev                   \\
                                                         & firmware                  & LPHD.PhyNam.swRev                   \\ \midrule
                \multirow{2}{*}{properties.serialNumber} & device                    & LPHD.PhyNam.serNum                  \\
                                                         & firmware                  & N / A                               \\ \bottomrule
            \end{tabular}%
            }
        \end{table}

    \subsection{Seamless integration with the IEC 61850 engineering process}
\begin{figure}[!htb]
            \centering
            \includegraphics[width=3in]{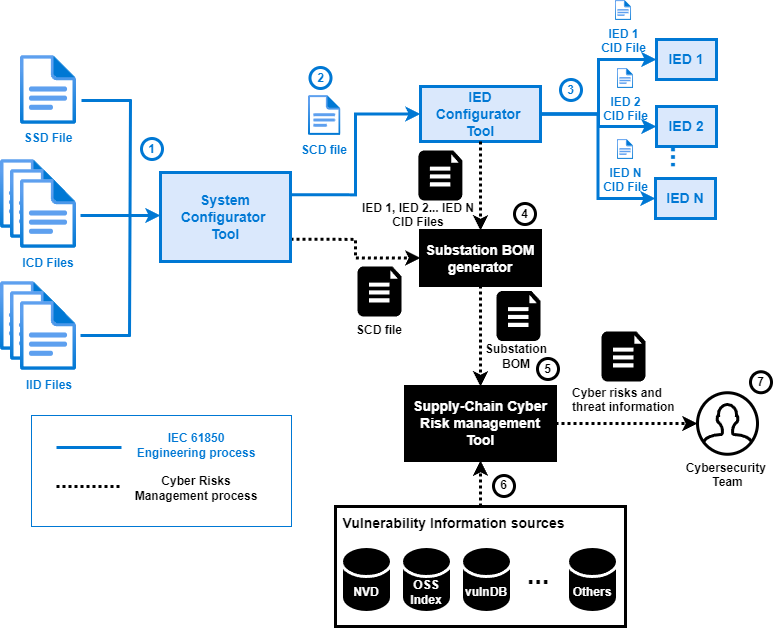}
            \caption{Flowchart of Substation BOM generation inside IEC 61850 engineering process}
            \label{fig:sbom_gen_flowchart}
        \end{figure}

        The proposed approach allows the integration of the Subs-BOM generation into the general engineering process of an IEC 61850 DS in Part 6 of the standard~\cite{iec_61850_part6}. Fig.~\ref{fig:sbom_gen_flowchart} shows the flowchart of the engineering process implementing this approach. By integrating the generation of the Subs-BOM into the engineering process, the effort to gather the list of assets is minimized. A comprehensive list of assets enhances the results of the management of cyber risks, enabling a more in-depth analysis. This asset inventory is critical to a quality management of the cyber risksº
        
        The proposed approach is as follows:

        \begin{enumerate}
            \item \textbf{Generation of the SCD File}: The System Configurator uses both the SSD and the ICD/IID files of the IEDs that will be deployed in the substation as input to generate the SCD file.

            \item \textbf{Generation of the CID Files}: The SCD file is then imported by IED configuration tools to generate the CID files containing the configuration instances for each of substation's IEDs.

            \item \textbf{Configuration of the IEDs}: The IED configurators configure each of the IEDs according to the generated CID files.

            \item \textbf{Generation of the Subs-BOM}: Upon completion of the IEC 61850 engineering process, that results in the correct configuration of substation IEDs, the generated SCD and CID files are used by the Subs-BOM generator to generate the Subs-BOM of the substation with the mandatory information. The Subs-BOM could be stored in the corresponding Supply-Chain Cyber risks management tool used by the utility.

            \item \textbf{Search in Available Data sources}: The Supply-Chain Cyber Risks Management tool correlates the information of the Subs-BOM with other Threat intelligence data sources, like vulnerability information repositories, to expand risk information of each of the assets deployed inside substation infrastructure.

            \item \textbf{Analysis of the Cyber Risk Information}: The generated cyber risk information is presented to the cybersecurity team to be studied and evaluated. The aim is to provide the cybersecurity team with the required information to allow the informed decision making around the actions to mitigate the identified risks.
        \end{enumerate}

    \subsection{Scope of Applicability and Model Preconditions}
        The Subs-BOM Schema needs a minimal amount of information to be properly built, and to extract it maximum potential. Nevertheless, in reality, there are some aspects that have to be considered:

        \begin{enumerate}
            \item \textbf{The IEC 61850 standard is an standard of minimums}: This means that only basic control and monitorization data and attributes are mandatory. This fact makes it difficult to standardize, and subsequently implement, engineering processes in multi-vendor environments. The proposed Subs-BOM schema requires that the various substation configuration SCD files contain the necessary data (\textit{e.g.},device nameplate attributes of each deployed device) to build it.

            \item \textbf{Modeling auxiliary equipment}: In an IEC 61850 DS usually exist other auxiliary equipment like firewalls or SCADA PCs that could not be modeled into the resulting SCD files, although they require an active management of their cybersecurity. The inclusion of this devices into the proposed Subs-BOM schema would require to manually include them, because of the requirements set by the IEC 61850 standard model.

            \item \textbf{Inclusion of risk-related information}: The IEC 61850 standard was not designed with cybersecurity in mind. Because of this, the SCD files do not include critical information required to perform risk-based decisions regarding to the supply chain. This is the case of the Common Platform Enumeration (CPE)~\cite{CPE_main} identifiers. To overcome this limitation, the CPE can be included into the SCD file by defining a private attribute. Private attribute definition is supported by IEC 61850 standard.    
            
            \item \textbf{CPE Identification}: Not all IED vendors register the CPE IDs into the CPE Product Dictionary maintained by the NIST. Therefore, some products require defining \textit{ad hoc} CPE identifiers.
        \end{enumerate}
\section{Experimental Setup}
\label{experimental_setup}
    DS are not isolated infrastructures, and in reality, DSOs have to manage and monitor multiple of them at the same time. This includes cybersecurity-related task, such as managing cyber-risks. For this purpose, the Subs-BOM schema is specially useful, by allowing DSOs to supervise and evaluate the cybersecurity status of several DS. The information obtained from the Subs-BOM can be use to take informed decisions about patching activities, and prioritizing by criticality.
    In this section, the proposed Subs-BOM schema is validated against a laboratory DS.

    \subsection{Description of the Experimental Setup}
        All the validation of the Subs-BOM was performed in the Smart Grid Cybersecurity Laboratory of TECNALIA. This laboratory allows to carry out tests and experiments over real assets, including all the electronic equipment of an IEC 61850 primary substation.

\begin{figure}[!htb]
            \centering
            \includegraphics[width=3in]{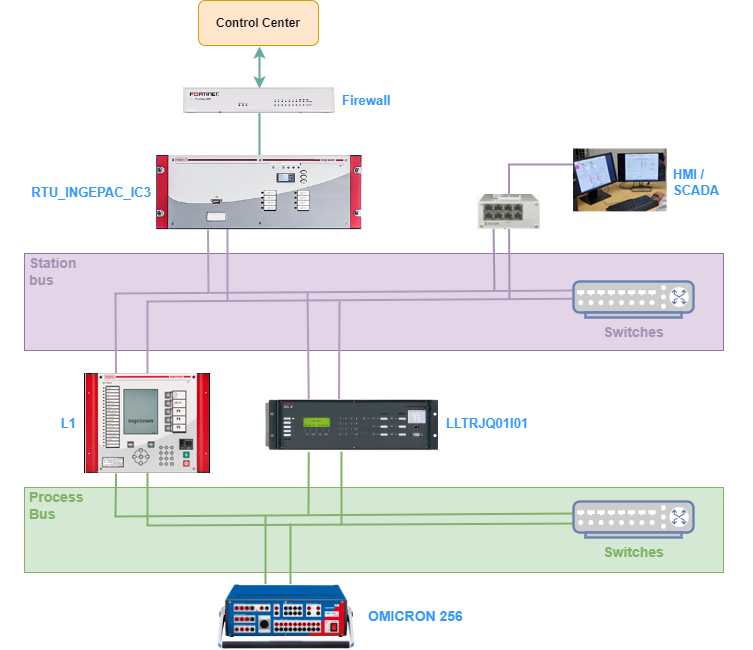}
            \caption{Laboratory IEC61850 communications architecture.}
            \label{fig:lab_comms_arch}
        \end{figure}

        
        The communication architecture of the substation of the laboratory is depicted in Fig.~\ref{fig:lab_comms_arch}, together with the existing IEC 61850 devices. To simulate the electric behavior of the substation, an Omicron 256 testing devices is used. It generates and injects IEC 61850 Sample value messages into the process bus simulating the values measured from electric equipment. The architecture of the laboratory substation is modeled within an SCD file according to Part 6 of the IEC 61850 standard.
        

        Finally, the experimental setup also includes the Dependency-Track software~\cite{dependency-track} by OWASP. This tool will serve as the Supply-Chain Cyber Risk Management Tool described in Fig.~\ref{fig:sbom_gen_flowchart}. This software is capable of importing the resulting Subs-BOM and analyzing its content. The Dependency-Track software is deployed in a virtualization platform in the laboratory.

    \subsection{Validation Process}
        The validation of the Subs-BOM is performed using the Dependency-Track software. By using an Open Source tool, we make sure that the proposed schema is recognized by CycloneDX-compatible tools. This process will cover the following points:
        \begin{itemize}
            \item \textbf{The format of the Subs-BOM file}: The validation has to check that the Dependency-Track software is capable of detecting the format of the proposed Subs-BOM schema.
            \item \textbf{Dependency extraction from the Subs-BOM}: The Dependency-Track software has to detect the dependencies defined in the Subs-BOM between the IEDs and their firmware.
            \item \textbf{Vulnerability analysis}: Finally, the Dependency-Track software has to detect potential vulnerabilities in public databases, and link them to the assets defined in the Subs-BOM.
        \end{itemize}
\section{Results and discussions}
\label{sec:resultsDiscussion}
    This section describes the results obtained in the validation of the proposed Subs-BOM schema.

    \subsection{Generation of the Subs-BOM}
        The SCD file representing the DB depicted in Fig. \ref{fig:lab_comms_arch} was parsed to extract the information relevant to the generation of its corresponding Subs-BOM. Table \ref{tab:mapping_useCase} shows the corresponding values for the RTU\_INGEPAC\_IC3 and the LLTRJQ01I01 IED.
        
        \begin{table*}[!htb]
            \centering
            \caption{Mapping of the SDC fields to the Subs-BOM for the laboratory DS.}
            \label{tab:mapping_useCase}
            \resizebox{\textwidth}{!}{%
            \begin{tabular}{@{}lllll@{}}
                \toprule
                \textbf{Subs-BOM ATTRIBUTE}              & \textbf{CycloneDX TYPE}   & \textbf{IEC 61850 ATTRIBUTE}        & \textbf{USE CASE IED 1}         & \textbf{USE CASE IED 2}               \\ \midrule
                \multirow{4}{*}{name}                    & device                    & IED.name                            & RTU\_INGEPAC\_IC3         & LLTRJQ01I01                     \\ \cmidrule(l){2-5} 
                                                         & \multirow{3}{*}{firmware} & LPHD.PhyNam.vendor                  & Ingeteam                  & ZIV Automation                  \\
                                                         &                           & LPHD.PhyNam.model                   & INGEPAC EF MD3 FC4140     & LLTRJQ01I01                     \\
                                                         &                           & LPHD.PhyNam.swRev                   & 8.1.0.20                  & irv8                            \\ \midrule
                \multirow{2}{*}{manufacturer}            & device                    & \multirow{2}{*}{LPHD.PhyNam.vendor} & \multirow{2}{*}{Ingeteam} & \multirow{2}{*}{ZIV Automation} \\
                                                         & firmware                  &                                     &                           &                                 \\ \midrule
                \multirow{2}{*}{version}                 & device                    & LPHD.PhyNam.hwRev                   & ingepac\_ef\_md             & unknown                         \\
                                                         & firmware                  & LPHD.PhyNam.swRev                   & 8.1.0.20                  & irv8                            \\ \midrule
                \multirow{2}{*}{properties.serialNumber} & device                    & LPHD.PhyNam.serNum                  & LA10821000001             & 142295                          \\
                                                         & firmware                  & N / A                               & N / A                     & N / A                           \\ \bottomrule
            \end{tabular}%
            }
        \end{table*}

    \subsection{Validation of the Subs-BOM Format}
        The Subs-BOM resulting from processing the SCD file of the laboratory was ingested into the Dependency-Track software. In this way, we validated that the format of the SBOM conforms to the CycloneDX v1.6 specification. As expected, the tool experiments a problem while parsing the Subs-BOM. This is because the new \textit{substation} component type that we have incorporated to the set of standardized component types is not supported by the tool. The tool does not warn of any other problem with the parsing of the Subs-BOM.
        
        The processing of the Subs-BOM allows obtaining cybersecurity information associated with the different components of the substation.

    \subsection{Assets and Dependencies Extraction from the Subs-BOM}
\begin{figure}[!htb]
            \centering
            \includegraphics[width=3in]{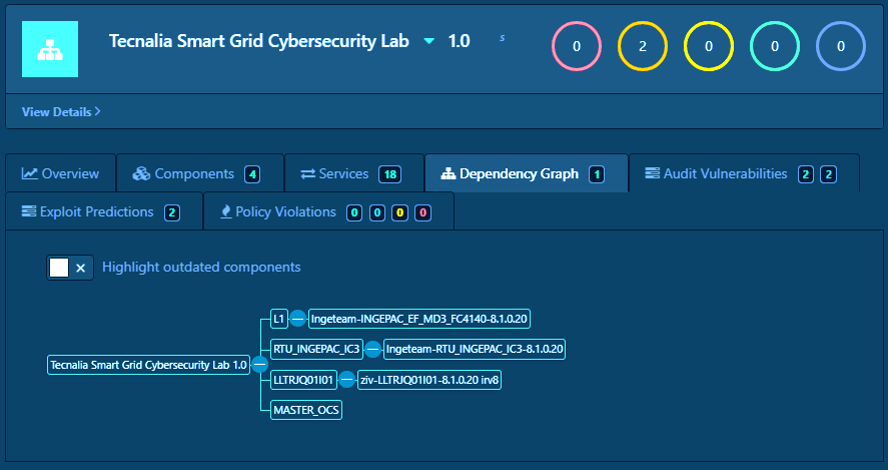}
            \caption{Dependency graph of Tecnalia's Smart Grid Cybersecurity Laboratory.}
            \label{fig:cyber_lab_depen_graph}
        \end{figure}

        The Dependency-Track software was capable of obtaining the asset list from the Subs-BOM (components according to CycloneDX naming convention). Moreover, the tool could also extract their dependencies, as shown in Fig.~\ref{fig:cyber_lab_depen_graph}.
        
    \subsection{Vulnerability Analysis} 
\begin{figure}[!htb]
            \centering
            \includegraphics[width=3in]{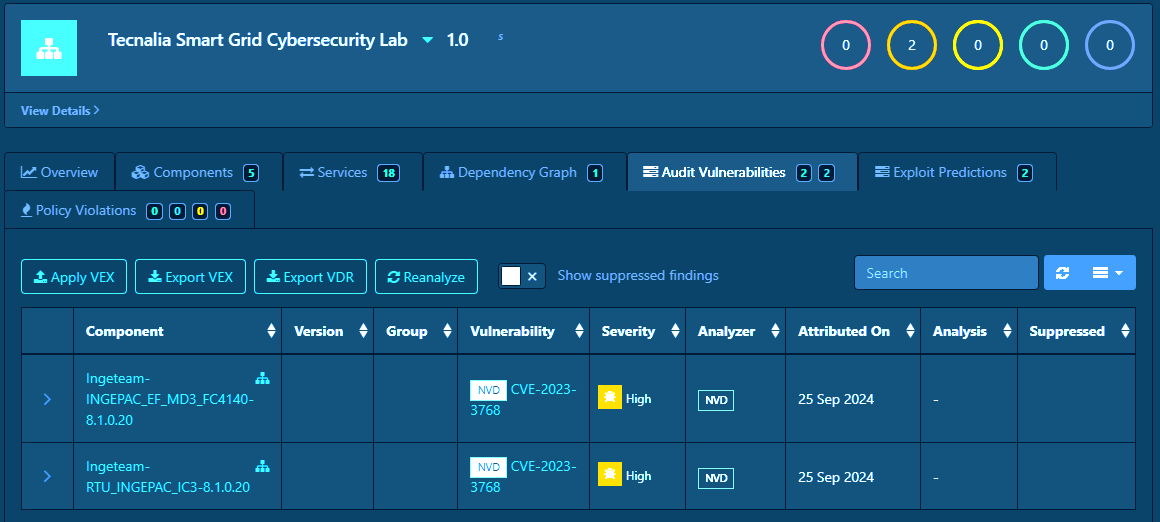}
            \caption{Vulnerabilities affecting components of Tecnalia's Smart Grid Cybersecurity Laboratory.}
            \label{fig:component_vulns}
        \end{figure}

        Based on information from the Subs-BOM, the Dependency-Track software could identify the vulnerabilities affecting the components deployed in the substation. Fig.~\ref{fig:component_vulns} shows that the two components for which their CPE is available are affected by the vulnerability CVE-2023-3768.
        
        Despite the tool indicating the existence of two vulnerabilities, in reality it is the same vulnerability. This is because this vulnerability is affecting the firmware that is installed in the devices. And therefore, due to the transitive property, the vulnerability also affects the device where the firmware is running. For this reason, CycloneDX requires to establish a relation with the corresponding firmware component for every device that is defined. The generated Subs-BOM has correctly captured this dependency.

\begin{figure}[!htb]
            \centering
            \includegraphics[width=3in]{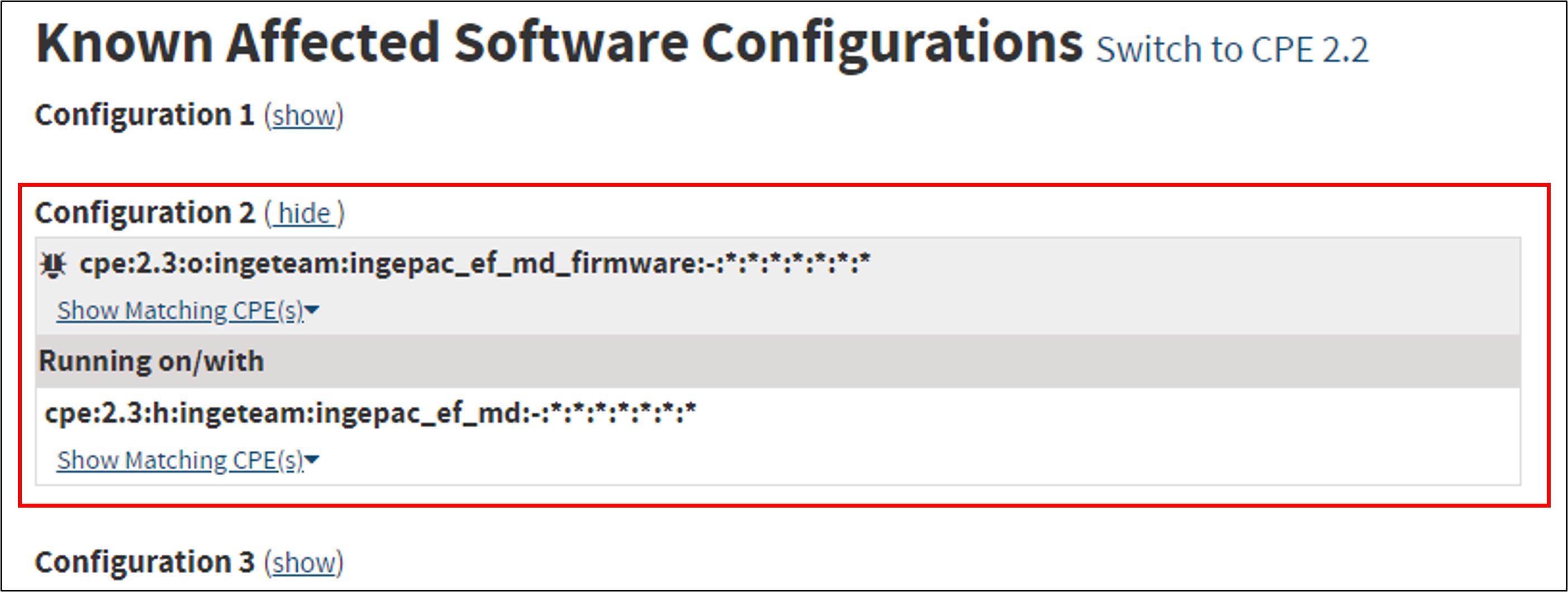}
            \caption{Information on components affected by vulnerability CVE-2023-3768 published by NIST.}
            \label{fig:affected_components}
        \end{figure}

        The information related to the vulnerabilities found in this analysis is shown in Fig.~\ref{fig:affected_components}. This data confirms the correctness of the information included in the Subs-BOM.
        
\section{Conclusion and Future Work}
\label{sec:conclusions}
    Smart grids have undergone a profound process of digitization where more and more connected substation are being deployed, following the IEC 61850 standard. Nevertheless, the IEC 61850 standard does not considers the cybersecurity aspects of DS. In this multi-vendor scenario, attackers have directed their efforts towards the device supply chain, rather than directly attacking DS.
    
    To manage the Supply Chain Cyber-Risks of IEC 61850 DS, we presented the Subs-BOM schema, which provides an accurate and complete inventory of the devices, the firmware they are running, and the services that are deployed into the DS, everything modeled from a cybersecurity perspective:

    \begin{itemize}
        \item It allows to manage a large number of devices.
        \item The Subs-BOM schema enhances the observability of IEDs.
        \item It allows to gather and update information about vulnerabilities in real time.
        \item The generation of the Subs-BOM can be automated and integrated in existing engineering processes.
    \end{itemize}

    We validated the Subs-BOM schema against the Dependency-Track software by OWASP. This validation proved that the schema is correctly recognized by CycloneDX-compatible tools. Moreover, the Dependency-Track software could track existing vulnerabilities to the IEDs represented by the Subs-BOM.

    As future work, the proposed Subs-BOM can be expanded to include information about the electric equipment that is controlled by the IEDs, and a standardized method to integrate the CPE identifier of each device. This would allow to measure the impact of a supply chain attack in a DS.
\bibliographystyle{ieeetr}
\bibliography{./bibliography}

\end{document}